\documentclass[12pt]{article}
\usepackage{amsxtra,amssymb,amsthm,amsmath,latexsym}

\theoremstyle{plain}

\numberwithin{equation}{section}

\newcommand{\opU}{\operatorname*{U}}

\newcommand{\refS}[1]{Section~\ref{S:#1}}

\def\R{{\mathbb R}}

\def\calN{{\mathcal N}}
\def\calR{{\mathcal R}}
\def\calS{{\mathcal S}}
\def\CalU{{\mathcal U}}
\def\oH{\buildrel\circ\over H}
\def\oH1{\buildrel\circ\over H\kern-.02in{}^1}

\def\tildebb{\widetilde\beta}
\def\tildegg{\widetilde\gamma}

\def\const{\hbox{\,const\,}}
\def\inf{\hbox{\,inf\,}}

\begin{document}
A.G. Ramm, Equations for the self-consistent field in random 
medium, Physics Letters A, 312, N3-4, (2003), 256-261.
%\begin{titlepage}
\title{                  
Equations for the self-consistent field in random medium}

\author{
A.G. Ramm\\
 Mathematics Department, Kansas State University, \\
 Manhattan, KS 66506-2602, USA\\
ramm@math.ksu.edu\\
http://www.math.ksu.edu/\,$\widetilde{\ }$\,ramm}

\date{}

\maketitle\thispagestyle{empty}

\begin{abstract}
\footnote{Math subject classification: 73D25, 73D50, 78A45; PACS 03.40Kf 
05.45.+b }
\footnote{key words: wave scattering, random media, small particles, 
acoustics, electromagnetics  }
An integral-differential equation is derived for the 
self-consistent (effective) field in the medium consisting of many 
small bodies randomly distributed in some region. 
Acoustic and electromagnetic fields are considered in such a 
medium. Each body has a characteristic dimension $a\ll\lambda$, 
where $\lambda$ is the wavelength in the free space.
 The minimal distance $d$ between any of the two bodies satisfies the 
condition $d\gg a$, but it may also satisfy the condition 
$d\ll\lambda$ in acoustic scattering.
In electromagnetic scattering our assumptions are $a\ll\lambda$
and $\lambda\ll d$.
 Using Ramm's theory of wave 
scattering 
by small bodies of arbitrary shapes, the author derives an 
integral-differential equation for the self-consistent acoustic or 
electromagnetic fields in the above medium.

\end{abstract}
%\end{titlepage}

\section{Introduction}
A general method to derive equations for the self-consistent
(effective) field in a medium consisting of many small particles is
proposed. The method is illustrated by the derivation of such equations
for acoustic and electromagnetic waves. The novel points in this paper 
include: 

1) the method for the derivation of the equations for the 
self-consistent field in a medium consisting of many small bodies
of arbitrary shapes; 

2) equations (2.29), (3.2)-(3.5),  and the formulas
for calculation of the polarizability tensors for bodies
of arbitrary shapes with the desired accuracy.

Equation (2.16) is of the type, obtained earlier in \cite{M}
by a different argument, and in \cite{R}. It is simpler than equation 
(2.29).
This can be easily explained physically: scattering of an acoustic wave
by small, in comparison with the wavelength in the free space, 
acoustically soft body is isotropic, and the scattered field
in the far-field zone is described by one scalar, the 
electrical capacitance of the perfect conductor of the same shape as the 
small body, and the scattered field is of order $O(a)$, where
 $a$ is the characteristic size of the small body.
If the small body is acoustically hard, that is, 
condition (2.20) holds on its boundary, then the scattering is 
anisotropic, the scattered field in  the far-field zone is described
by a tensor, and this field is of order $k^2a^3$, that is, much smaller 
(by a factor of order $O((ka)^2)$) than for the acoustically soft body of 
the same size $a$ and of the same shape. Here  $k$ is 
the wavenumber in the free space.

In \cite{B} the Lorentz-Lorenz formula is derived. This formula relates 
the 
polarizability of the uniform dielectric to the density of the 
distribution of molecules and the polarizability of these
molecules. In this theory one assumes that the polarizabilty
in the matter is a constant vector, the molecules are modeled as 
identical spheres uniformly distributed in the space.
In this case the polarizability tensor is proportional to the
unit matrix, and the coefficient of proportionality is the cube of the 
radius of the small sphere times some constant. This, together
with additional assumptions, yields a 
relation between the dielectric constant of the medium, the 
polarizability of the molecule, and the number of the molecules per unit 
volume. The derivation of this formula in \cite{B} is based on 
the equation of electrostatics.

Our basic physical assumptions, (1.1), allow for rarefied medium, when
$d>>\lambda$, but also, in acoustic wave scattering theory, for medium 
which is dense, when 
$a<<d<<\lambda$. Equations (2.29) and (3.2)-(3.5), that we have derived,
have an unusual feature: the integrand depends on the direction from $y$ 
to $x$. This happens because of the anisotropy of the scattered field
in the case of non-spherical homogeneous small bodies.

A possible application of equation (2.16) is a method for finding the 
density of the distribution of small bodies from the scattering data.
Namely, the function $C(y)$ in (2.16)-(2.17) determines this distribution.
On the other hand, this function can be determined from 
the field scattered by the region $\calR$. The uniqueness results and 
computational methods for solving this inverse scattering problem
are developed in \cite{R1},\cite {R2}, and \cite {R3}
(see also the book \cite{R4} published in 2005).

Below we study the dynamic fields, so that the wavenumber is positive.

 Consider a random medium consisting of many small bodies $D_j$, 
$1\leq j\leq J$, $J\sim 10^{23}$, located in a region $\calR$. Let $a_j$ 
be the radius of the body $D_j$, defined as 
$a_j:=\frac{1}{2} \underset{x,y\in D_j}{\max}|x-y|$, and 
$a:=\underset{1\leq j\leq J}{\max} a_j$. 
We assume $a\ll\lambda$, where $\lambda$ is the wavelength of the field in the free space (or in a homogeneous space in which the small bodies are embedded). Let 
$d:=\underset{x\in D_j,y\in D_j, 1\leq i, j\leq J, i\not= j}{\min\qquad\qquad}  |x-y| $.
Assume
\begin{equation}\label{e1.1}
 a\ll \lambda, \ a\ll d.
 \end{equation}
We do not assume that $d\gg \lambda$ in the acoustic wave scattering,
but assume this in electromagnetic wave scattering. The difference in the
physical assumptions between acoustic and electromagnetic theory
is caused by the necessity to apply twice the operation $\nabla \times$ to 
the expression of the type $p\frac {e^{ikr}}{r}$
in the electromagnetic theory, where $p$ is a vector independent of $r$
and $r$ is the distance from a small body to the observation point.

We consider acoustic field in the above medium, and derive in \refS{2} an 
integral-differential equation for the self-consistent field in this medium. 
The notion of the self-consistent field is defined in \refS{2}. 
Roughly speaking, it is the field, acting on one of the small bodies 
from all other bodies, plus the incident field.

In \refS{3} we derive a similar equation for the self-consistent 
electromagnetic field in the medium.
% and the formulas for the tensors 
%$\varepsilon_{ij}(x)$, $\mu_{ij}(x)$ and $\sigma_{ij}(x)$ 
%of the electrical permeability, magnetic permeability and conductivity in 
%the above random medium.

Each small body may have an arbitrary shape. The key results from 
 \cite{R}, that we use, are the formulas for the 
$S\hbox{-matrix}$ for acoustic and electromagnetic wave scattering 
by a single homogeneous small body of an arbitrary shape. 
These formulas are given in \refS{2} and \refS{3}.
Wave propagation in random media is studied in \cite{I}.

\section{Acoustic fields in random medium.}\label{S:2}
Assume first that the small bodies are acoustically soft, that is $u\mid_{S_j}=0$, where $S_j$ is the boundary of $D_j$. The governing equation for the acoustic pressure $u$ is 
\begin{equation}\label{e2.1}
  (\nabla^2+k^2)u=0 \hbox{\ in\ } D':=\R^3\backslash D,  
  \qquad D:=\opU^J_{j=1} D_j,
  \end{equation}
\begin{equation}\label{e2.2}
  u=0 \hbox{\ on\ } S:=\opU^J_{j=1} S_j,
  \end{equation}
\begin{equation}\label{e2.3}
  u=u_0+v, \qquad u_0:=e^{ik\nu \cdot x}, \qquad \nu\in S^2,
  \end{equation}
\begin{equation}\label{e2.4}
  r \left( \frac{\partial v}{\partial r} -ikv \right) \to 0,   \qquad r:=|x|,
  \end{equation}
where the direction $\nu\in S^2$ of the incident field is given, 
$k=\frac{2\pi}{\lambda}=\const>0$ is the wave number, $S^2$ is the unit 
sphere in $\R^3$.

Let us look for $v$ of the form:
\begin{equation}\label{e2.5}
  v(x)=\sum^J_{j=1} \int_{S_j} g(x,s)\sigma_j(s) ds,
  \qquad g(x,y):= \frac{e^{ik|x-y|}}{4\pi|x-y|}.
  \end{equation}
If \eqref{e1.1} holds, and $\underset{1\leq i\leq J}{\min} |x-s_j|\gg a$, 
$s_j\in S_j$, then \eqref{e2.5} can be written as
\begin{equation}\label{e2.6}
  v(x)= \sum^J_{j=1} g(x,s_j) Q_j[1+O(ka)],
  \qquad Q_j:=\int_{S_j} \sigma_j ds,
  \end{equation}

{\it 
Define the self-consistent field $u_e$ at the point $s_m\in S_m$ by the formula
\begin{equation}\label{e2.7}
  u_e(s_m)=u_0(s_m) +\sum^J_{j=1,j\not= m} g(s_m,s_j) Q_j,
  \end{equation}
and at any point $x$, such that 
\begin{equation}\label{e2.8}
  \inf_j|x-s_j| \gg a,
  \end{equation}
by the formula:
\begin{equation}\label{e2.9}
  u_e(x)=u_0(x)+\sum^J_{j=1} g(x,s_j)Q_j.
  \end{equation}
}
If \eqref{e2.8} holds, then 
$$ \bigg| \sum^J_{j=1} g(x,s_j) Q_j 
  - \sum_{j=1, j\not= m} g(x,s_j) Q_j \bigg|    \ll |u_e(x)|,$$
that is, removal of one small body does not change the self-consistent field in the region which contains no immediate neighborhood of this body. On the surface $S$ of the body the total field $u=0$, so the self-consistent field on $S_m$ differs from $u$, while at a point $x$ such that \eqref{e2.8} holds, $u(x)=u_e(x)+o(1)$, if \eqref{e1.1} holds.

Let us derive a formula for $Q_j$. By \eqref{e2.2} one gets:
\begin{equation}\label{e2.10}
  0=u_e(s_m) + \int_{S_m} \frac{ \sigma_m (s)ds}{4\pi|s_m-s|} 
+\varepsilon_m,
  \end{equation}
where
\begin{equation}\label{e2.11}
  \varepsilon_m:=\int_{S_m} [g(s_m,s) - g_0(s_m,s)] \sigma_mds,
  \qquad g_0:=\frac{1}{4\pi|s_m-s|},
  \end{equation}
so that 
\begin{equation}\label{e2.12} 
  \underset{m}{\max} |\varepsilon_m|=o(1) \hbox{\ as\ } ka\to 0.
  \end{equation}
Thus, one may neglect $\varepsilon_m$ in \eqref{e2.10} and consider the resulting equation
\begin{equation}\label{e2.13}
  \int_{S_m} \frac{\sigma_m ds}{4\pi|s_m-s|} = -u_e(s_m)
  \end{equation}
as an equation for the charge distribution $\sigma_m$ on the surface $S_m$ of a perfect conductor charged to the potential $-u_e(s_m)$.
Then, by \eqref{e2.6},
\begin{equation}\label{e2.14}
  Q_m:= \int_{S_m} \sigma_m ds = -C_m u_e(s_m),
  \end{equation}
where $C_m$ is the electrostatic capacitance of the conductor $D_m$.
From \eqref{e2.9} and \eqref{e2.14} one gets
\begin{equation}\label{e2.15}
  u_e(x)=u_0(x) -\sum^J_{j=1} g(x,s_j) C_j u_j(s_j)  +o(1),
  \end{equation}
as $ka\to 0$. 

Let us emphasize the physical assumptions we have used in the derivation 
of (2.15). First, the assumption $ka<<1$ allows one to claim that,
uniformly with respect to all small bodies, 
the term $o(1)$ in (2.15) tends to zero
as $ka\to 0$. Secondly, the assumption $d>>a$ allows one to claim that
the $m-$th small body is in the far zone with respect to the $j-$th body
for any $j\neq m$. The expression under the sum in (2.15) is the 
field, scattered by $j-$th body and calculated at the point $x$, such
that $|x-s_j|>>a$, that is, in the far zone from the $j-$th body.
So, physically, the equatipons for the self-consistent field in the 
medium, derived in this paper, are valid not only for the rarified medium
(that is, when $a<<\lambda$ and $d>>\lambda$, but also for not too dense
medium, that is, when $d>>a$ and $a<<\lambda$, but, possibly, 
$d<<\lambda$. 
  
The limiting equation for $u_e(x)$ is:
\begin{equation}\label{e2.16}
  u_e(x) = u_0(x) - \int_\calR g(x,y) C(y) u_e(y) dy,
  \end{equation}
where
\begin{equation}\label{e2.17}
  C(y)dy=\sum_{s_j\in dy} C_j,
  \end{equation}
and the summation is taken over all small bodies located in the volume $dy$ around point $y$.
If one assumes that the capacitances $C_j$ are the same for all these 
bodies around point $y$, and are equal to $c(y)$, then $C(y)=c(y) N(y)$, 
where $N(y)$ is the number of small bodies in the volume $dy$.

Equation  \eqref{e2.16} is the integral equation for the self-consistent 
field in the medium in the region $\calR$. This field satisfies the 
Schr\"odinger equation:
\begin{equation}\label{e2.18}
  [\nabla^2+k^2-q(x)] u_e=0,  \qquad q(x):= C(x).
  \end{equation}
Since $C_j\sim a $ in \eqref{e2.17}, and the number $N$ of the terms in 
the sum \eqref{e2.17} is $N=O(\frac{1}{d^3})$, provided that $dy$ is a 
unit cube, one concludes that $\frac{a}{d^3}=O(1)$, so
\begin{equation}\label{e2.19}
  d=O \left( a^{\frac{1}{3}}\right).
  \end{equation}
If one had $N a^3=O(1)$, i.e., small bodies have nonzero limit of 
volume density, then {\it the assumption $d\gg a$ would be violated.}

Let us now assume that the small bodies are acoustically hard, i.e., the 
Neumann boundary condition
\begin{equation}\label{e2.20}
  \frac{\partial u}{\partial \calN}=0 \hbox{\ on\ } S,
  \end{equation}
replaces \eqref{e2.2}, $\calN$ is the outer normal to $S$. 
In this case the derivation of the equation for $u_e(x)$ is more complicated, 
because the formula for $Q_j$ is less simple. It is proved in 
\cite{R} that for the boundary condition \eqref{e2.20} one gets
\begin{equation}\label{e2.21}
  Q_j=ikV_j \beta^{(j)}_{pq} n_p \frac{\partial u_e}{\partial x_q} + 
V_j \Delta u_e.
  \end{equation}
Here and below, one {\it sums up over the repeated indices}, $V_j$ is the 
volume 
of $D_j$, $\Delta u_e$ is the Laplacean, $n_p:= \frac{x_p-y_p}{|x-y|}$, 
the small body $D_j$ is located around point $y$, the scattered field is 
calculated at point $x$, $\beta^{(j)}_{pq}$ is the magnetic polarizability 
tensor of $D_j$ which is defined by the formula \cite{R}
\begin{equation}\label{e2.22}
  \beta_{pq}:= \alpha_{pq}(-1),
  \end{equation}
and $\alpha_{pq}(\gamma)$, $\gamma:=\frac{\varepsilon-\varepsilon_0}
{\varepsilon  +\varepsilon_0}$, is the electric polarizability tensor, defined by the formula:
\begin{equation}\label{e2.23}
  P_p=\alpha_{pq}(\gamma) V_q \varepsilon_0 E_q.
  \end{equation}
Here $P$ is the dipole moment induced on the dielectric body $D_j$, with the 
dielectric constant $\varepsilon$, placed in the electrostatic field $E$ 
in the homogeneous medium with the dielectric constant $\varepsilon_0$.

Analytical formulas for calculation of $\alpha_{pq}(\gamma)$ with an 
arbitrary accuracy, in terms of the geometry of $S_j$, are derived in 
\cite{R}.
These formulas are:
\begin{equation}\label{e2.24}
  \alpha^{(n)}_{pq} :=\frac{2}{V} \sum^n_{m=0} 
  \left( \frac{-1}{2\pi} \right) \frac{\gamma^{n+2} - \gamma^{m+1}}{\gamma-1} 
  b^{(m)}_{pq}, \qquad n\geq 1,
  \end{equation}
where
\begin{equation}\label{e2.25}
  b^{(0)}_{pq} := V\delta_{pq}, 
  \  b^{(1)}_{pq} := \int_{S_j} \int_{S_j} \frac{\calN_p(s) \calN_q(t) dsdt}
{r_{st}},  \ r_{st}:=|s-t|, 
  \delta_{pq}=\begin{cases}1, &\hbox{if $p=q$},\\ 0, &\hbox{if 
$p\not=q$},\end{cases}
  \end{equation}
\begin{equation}\label{e2.26}
  \begin{aligned}
  b^{(m)}_{pq} 
&
  := \int_{S_j} \int_{S_j} ds dt \calN_p(s)\calN_q(t)
  \underbrace{\int_{S_j} \dots \int_{ S_j}}_{m-1}
  \frac{1}{r_{st}} \psi(t_1,t)\dots 
  \ \psi(t_{m-1}, t_{m-2}) dt_1 \dots dt_{m-1},
\\
  & \qquad  
  \ \psi(t,s):=\frac{\partial}{\partial \calN_t} \frac{1}{r_{st}},\quad
  \alpha^{(1)}_{pq} (\gamma) = 2 (\gamma+\gamma^2) \delta_{pq} 
   - \frac{\gamma^2    b^{(1)}_{pq}}{\pi V},
  \end{aligned}
\end{equation}
\begin{equation}\label{e2.27}
  \beta^{(1)}_{pq} = -\frac{b^{(1)}_{pq}}{\pi V},
  \end{equation}
and
\begin{equation}\label{e2.28}
  | \alpha_{pq}(\gamma) -\alpha^{(n)}_{pq}(\gamma) | =O(q^n),
  \qquad 0<q<1,
  \end{equation}
that is, \eqref{e2.24} gives a convergent approximation of the tensor $\alpha_{pq}(\gamma)$.

From \eqref{e2.9} and \eqref{e2.21} one gets
\begin{equation}\label{e2.29}
  u_e(x)=u_0(x) + \int_\calR g(x,y) 
  \left[ ik \beta_{pq}(y)\frac{x_p-y_p}{|x-y|} 
     \frac{\partial u_e(y)}{\partial y_q}  + \Delta u_e(y) \right]
  V(y)dy,
  \end{equation}
where $V(y)$ and $\beta_{pq}(y)$ are defined by the formulas
\begin{equation}\label{e2.30}
  V(y)dy=\sum_{s_j\in dy} V_j,\quad
  \beta_{pq}(y)V(y) dy =\sum_{s_j\in dy} \beta_{pq,j}V_j, 
  \end{equation}
where $\beta_{pq,j}$ is the magnetic polarizability tensor of the $j-$th 
small body, 
and the summation is over all small bodies in the volume $dy$ around point 
$y$, so that $V(y)$ is the density of the distribution of the 
volumes of small bodies at a 
point $y$.

Equation (2.29) is approximate, in contrast to (2.16).
Indeed, if $a<<d$ then $\lim_{\frac a d \to 0}\sum_{s_j \in dy}V_j=0$
because $V_j=O(a^3)$ and the number of small bodies in the unit volume
is $O(\frac 1 {d^3})$, so  $\lim_{\frac a d \to 0}\sum_{s_j \in dy}V_j=
O(\frac {a^3}{d^3} dy)=0$. However, if the Dirichlet condition  
(2.2) holds, rather than the Neumann one  (2.20),
then $\frac a {d^3}=O(1)$ and equation  (2.16) is exact in the 
limit $\frac a d \to 0$ with $C(y)$ defined in (2.17).

The novel feature of equation \eqref{e2.29} is the dependence of the
integrand in \eqref{e2.29} on the direction $x-y$. This one can
understand, if one knows that the acoustic wave scattering by a small soft
body $D_j$ is isotropic and depends on one scalar, the electrostatic
capacitance of the conductor $D_j$, while acoustic wave scattering by a
small hard obstacle is anisotropic and depends on the tensor
$\beta^{(j)}_{pq}$.

Finally, if the third boundary condition holds:
\begin{equation}\label{e2.31}
  \frac{\partial u}{\partial \calN} +hu=0 \hbox{\ on\ } S
  \end{equation}
then (see \cite{R}), if $h$ is not too small, one has:
\begin{equation}\label{e2.32}
  Q_j= -\frac{h|S_j|}{1+h|S_j|C^{-1}_j} u_e,
  \end{equation}
where $|S_j|$ is the area of $S_j$, so that
\begin{equation}\label{e2.33}
    u_e(x) =u_0(x) -\int_\calR g(x,y) b(y)u_e(y)dy,
  \end{equation}
where
\begin{equation}\label{e2.34}
  b(y)dy =\sum_{s_j\in dy} \frac{h|S_j|}{1+h|S_j|C^{-1}}.
  \end{equation}

\section{Electromagnetic waves in random medium}\label{S:3}

 In this section our basic assumptions are:
\begin{equation}\label{e3.0}
 a<<\lambda \quad d>>\lambda.
\end{equation}
The reason for the change in the assumption compared with (1.1),
where $d$ is not necessarily greater than $\lambda$,  is
the following: formula (3.4) below is valid if $a<<\lambda$ 
and $d>>\lambda$,
where  $d$ is the distance from a small body to the point of observation.
This comes from applying twice the operation of $\nabla \times$ to the 
vector potential (see, e.g., \cite{R}, p.98-99). 
In acoustic scattering the formula for the scattered field (see e.g. 
(2.7)) is valid if $d>>a$, while formula (3.4) (see below) is valid for 
$d>>\lambda$
and $a<<\lambda$.
 
Let $\CalU:= \binom{E}{H}$. Denote by $\calS$ the 6x6 matrix which sends 
$\CalU$ 
into $g \CalU_{sc}$, where  $g \CalU_{sc}$ is the scattered field,
 $g:=\frac{e^{ikr}}{r}$, $r$ is the distance 
from a small body located at a point $y$ to the observation point $x$, 
$r=|x-y|$.

If $\calS$ is known, then the equation for the self-consistent field 
$ \CalU_e$ in the random medium situated in a region $\calR$, and 
consisting of many small bodies, satisfying conditions (3.1), is 
\begin{equation}\label{e3.1}
   \CalU_e=\CalU_0+\int_\calR g(x,y) \calS \left( y,\frac{x-y}{r} \right) 
 \CalU_e(y)dy,
  \end{equation}
where
\begin{equation}\label{e3.2}
 \calS \left( y,\frac{x-y}{r} \right) dy 
  = \sum_{s_j\in dy} \calS^{(j)} \left(s_j, \frac{x-s_j}{|x-s_j|} \right)
  \end{equation}
as follows from the argument given for the derivation of \eqref{e2.29}.

Let us give a formula for $\calS^{(j)}$, assuming without loss of generality, 
that the origin is situated inside a single body 
$D_j$, which has 
parameters $\varepsilon_j$, $\mu_j$, and $\sigma_j$, (dielectric 
permittivity, magnetic permeability, and conductivity, respectively), and 
drop index $j$ in the formula for $\calS^{(j)}$.

In \cite{R} one can find the formulas
\begin{equation}\label{e3.3}
  E_{sc}(\nu') = \frac{k^2}{4\pi}
  \left( \frac{1}{\varepsilon_0} [\nu',[P,\nu']] + 
\sqrt{\frac{\mu_0}{\varepsilon_0}} [M,\nu'] \right),
  \end{equation}
 \begin{equation}\label{e3.4}
    H_{sc}(\nu')= \sqrt{\frac{\varepsilon_0}{\mu_0}} [\nu',E_{sc}(\nu')] =
\frac{k^2}{4\pi}\left( 
\frac{1}{\sqrt{\varepsilon_0 \mu_0}} [\nu',P] +[\nu',[M,\nu']] \right),
  \end{equation}
where $\nu'$ is the unit vector in the direction of the scattered wave.

Formulas \eqref{e3.3}, \eqref{e3.4} can be written as
\begin{equation}\label{e3.5}
  \CalU_{sc} =\calS\CalU=\frac{k^2V}{4\pi} 
  \begin{pmatrix} 
    \alpha-\nu'(\nu',\alpha\cdot), & -(\mu^3_0 
\varepsilon^{-1}_0)^{\frac{1}{2}} 
[\nu',\tildebb\cdot] \\
    (\frac{\varepsilon_0} {\mu_0})^{\frac{1}{2}} [\nu',\alpha\cdot], & 
\mu_0(\tildebb-\nu'(\nu',\tildebb\cdot))
  \end{pmatrix}
  \binom{E}{H}
  \end{equation}
where the matrix $\calS$ is expressed in terms of the tensors 
$\alpha:=\alpha_{pq}$ and $\beta:=\beta_{pq}$, because $P$ is calculated 
in 
\eqref{e2.23},
\begin{equation}\label{e3.6}
  M_i=\alpha_{ij}(\tildegg) V \mu_0 H_j + \beta_{ij} V\mu_0 
H_j:=\tildebb_{ij} V \mu_0 H_j,\quad \tildegg 
  :=\frac{\mu-\mu_0}{\mu+\mu_0},
  \quad \tildebb_{ij}:= \alpha_{ij}(\tildegg)+\beta_{ij}.
  \end{equation}
$\beta_{ij}$ is defined in \eqref{e2.22}, and 
$(\nu,\alpha\cdot)E:=(\nu,\alpha E)$, 
$[\nu,\alpha\cdot]:= [\nu,\alpha E]$.

\end{document}